\documentclass[conference,hidelinks]{IEEEtran}

\usepackage{graphicx,url}
\usepackage[english]{babel}
\addto\captionsenglish{}
\usepackage[utf8]{inputenc}
\usepackage{amsmath}
\usepackage{amssymb}
\usepackage{lipsum}
\usepackage{mathtools}
\usepackage{cuted}
\usepackage{cite}
\usepackage{graphicx}
\usepackage{balance}
\usepackage{tikz}
\usepackage{pgfplots}
\usepackage{tikzscale} 
\pgfplotsset{compat=newest} 
\usepackage{scalefnt}
\usepackage{orcidlink}
\usepackage{dblfloatfix}
\usepackage[acronym,shortcuts]{glossaries}
\usepackage{algorithm, algpseudocode}
\usepackage{algcompatible}
\usepackage{bm}

\newacronym{RPE}{RPE}{radar parameter estimation}
\newacronym{OTFS}{OTFS}{orthogonal time frequency space}
\newacronym{AFDM}{AFDM}{affine frequency division multiplexing}
\newacronym{MIMO}{MIMO}{multiple-input multiple-output}
\newacronym{SISO}{SISO}{single-input single-output}
\newacronym{ISAC}{ISAC}{integrated sensing and communications}
\newacronym{3D}{3D}{three-dimensional}
\newacronym{2D}{2D}{two-dimensional}
\newacronym{1D}{1D}{one-dimensional}
\newacronym{RX}{RX}{receiver}
\newacronym{TX}{TX}{transmitter}
\newacronym{BF}{BF}{beamforming}
\newacronym{mmWave}{mmWave}{millimeter-wave}
\newacronym{SotA}{SotA}{state-of-the-art}
\newacronym{ULA}{ULA}{uniform linear array}
\newacronym{QAM}{QAM}{quadrature amplitude modulation}
\newacronym{ISFFT}{ISFFT}{inverse symplectic finite Fourier transform}
\newacronym{SFFT}{SFFT}{symplectic finite Fourier transform}
\newacronym{AWGN}{AWGN}{additive white Gaussian noise}
\newacronym{OFDM}{OFDM}{orthogonal frequency division multiplexing}
\newacronym{OCDM}{OCDM}{orthogonal chirp division multiplexing}
\newacronym{BS}{BS}{base station}
\newacronym{UE}{UE}{user equipment}
\newacronym{DFT}{DFT}{discrete Fourier transform}
\newacronym{IDFT}{IDFT}{inverse discrete Fourier transform}
\newacronym{IFFT}{IFFT}{inverse fast Fourier transform}
\newacronym{TD}{TD}{time-domain}
\newacronym{wlg}{wlg}{without loss of generality}
\newacronym{CP}{CP}{cyclic prefix}
\newacronym{DAFT}{DAFT}{discrete affine Fourier transform}
\newacronym{DAF}{DAF}{discrete affine Fourier}
\newacronym{IDAFT}{IDAFT}{inverse discrete affine Fourier transform}
\newacronym{CPP}{CPP}{\textit{chirp-periodic} prefix}
\newacronym{IDZT}{IDZT}{inverse discrete Zak transform}
\newacronym{DZT}{DZT}{discrete Zak transform}
\newacronym{ICI}{ICI}{inter-carrier interference}
\newacronym{BER}{BER}{bit error rate}
\newacronym{DoF}{DoF}{degrees-of-freedom}
\newacronym{FD}{FD}{full-duplex}
\newacronym{SIMO}{SIMO}{single-input multiple-output}
\newacronym{MISO}{MISO}{multiple-input single-output}
\newacronym{AoD}{AoD}{angle-of-departure}
\newacronym{AoA}{AoA}{angle-of-arrival}
\newacronym{RF}{RF}{radio frequency}
\newacronym{SIM}{SIM}{stacked intelligent metasurfaces}
\newacronym{FPGA}{FPGA}{field programmable gate array}
\newacronym{UPA}{UPA}{uniform planar array}
\newacronym{CC}{CC}{communication-centric}
\newacronym{I/O}{I/O}{input-output}
\newacronym{iid}{i.i.d.}{independent and identically distributed}
\newacronym{IoT}{IoT}{internet of things}
\newacronym{V2X}{V2X}{vehicle-to-everything}
\newacronym{NTN}{NTN}{non-terrestrial network}
\newacronym{LEO}{LEO}{low-earth orbit}
\newacronym{THz}{THz}{terahertz}
\newacronym{EM}{EM}{electromagnetic}
\newacronym{RIS}{RIS}{reconfigurable intelligent surface}
\newacronym{DoA}{DoA}{direction-of-arrival}
\newacronym{DD}{DD}{doubly-dispersive}
\newacronym{ODDM}{ODDM}{orthogonal delay-Doppler division multiplexing}
\newacronym{LoS}{LoS}{line-of-sight}
\newacronym{NLoS}{NLoS}{non-line-of-sight}
\newacronym{6G}{6G}{sixth generation}
\newacronym{MPDD}{MPDD}{metasurfaces-parametrized DD}
\newacronym{GaBP}{GaBP}{Gaussian Belief Propagation}
\newacronym{MSE}{MSE}{mean-squared-error}
\newacronym{sIC}{soft IC}{soft interference cancellation}
\newacronym{soft RG}{soft RG}{soft replica generation}
\newacronym{BG}{BG}{belief generation}
\newacronym{SGA}{SGA}{scalar Gaussian approximation}
\newacronym{CLT}{CLT}{central limit theorem}
\newacronym{PDF}{PDF}{probability density function}
\newacronym{QPSK}{QPSK}{quadrature phase-shift keying}
\newacronym{OQAM}{OQAM}{offset quadrature amplitude modulation}
\newacronym{LMMSE}{LMMSE}{linear minimum mean square error}
\newacronym{SNR}{SNR}{signal-to-noise ratio}
\newacronym{OOBE}{OOBE}{out-of-band emissions}
\newacronym{PAPR}{PAPR}{peak-to-average power ratio}
\newacronym{AFBM}{AFBM}{Affine Filter Bank Modulation}
\newacronym{FBMC}{FBMC}{Filter Bank Multicarrier Modulation}
\newacronym{PPN}{PPN}{polyphase network}
\newacronym{SIR}{SIR}{signal-to-interference ratio}

\begin{document}

%

%
%
\title{Low-Complexity Receiver Design for \\ Affine Filter Bank Modulation}

\author{\IEEEauthorblockN{Kuranage Roche Rayan Ranasinghe\textsuperscript{\orcidlink{0000-0002-6834-8877}}\IEEEauthorrefmark{2}, Bruno S. Chang\textsuperscript{\orcidlink{0000-0003-0232-7640}}\IEEEauthorrefmark{1} and Giuseppe Thadeu Freitas de Abreu\textsuperscript{\orcidlink{0000-0002-5018-8174}}\IEEEauthorrefmark{2}} \vspace{1ex}

\IEEEauthorblockA{
\IEEEauthorrefmark{2}School of Computer Science and Engineering, Constructor University, Bremen, Germany \\[0.15ex]
\IEEEauthorrefmark{1}CPGEI/Electronics Department, Federal University of Technology - Paraná, Curitiba, Brazil \\[0.15ex]
}
\vspace{-4ex}
}

\maketitle

\begin{abstract} 
We propose a low-complexity receiver structure for the recently introduced \ac{AFBM} scheme, which is a novel waveform designed for \ac{ISAC} systems operating in \ac{DD} channels.
The proposed receiver structure is based on the \ac{GaBP} framework, making use of only element-wise scalar operations to perform detection of the transmitted symbols.
Simulation results demonstrate that \ac{AFBM} in conjunction with \ac{GaBP} outperforms \ac{AFDM} in terms of \acp{BER} in \ac{DD} channels, while achieving very low \ac{OOBE} in high-mobility scenarios.
\end{abstract}

\begin{IEEEkeywords}
Doubly-dispersive, ISAC, AFDM, AFBM.
\end{IEEEkeywords}

\IEEEpeerreviewmaketitle

\glsresetall
\section{Introduction}

Wireless communications systems are increasingly required to support a wide range of applications  such as high-mobility communications \cite{Jingxian2016}, computing functionalities \cite{Ranasinghe_ICNC2025,RanasingheTWC2025ICC}, and \ac{ISAC} \cite{Zhiqing_IoT2023, NuriaISAC2024, RanasingheTWC2025}.
The high mobility and/or frequencies associated with such applications, however, imply the presence of \ac{DD} channels, which are known \cite{Svante2007} to lead to severe inter-carrier interference and consequent link performance degradation of to \ac{OFDM}, the waveform used in most modern communications systems.

To overcome this challenge, several novel waveforms have been proposed in the last years \cite{Rou_SPM_2024}, with \ac{AFDM} being one of the most recent and promising solutions.
The \ac{AFDM} \cite{bemani2023} waveform maps data symbols into chirp-based subcarriers in the \ac{DAF} domain, and is capable of attaining full diversity in \ac{DD} channels via the adjustment of chirp parameters, which are tuned based on the channel characteristics.
However, \ac{AFDM} suffers from a high \ac{PAPR} and poor spectral containment, leading to challenges in implementation such as inefficient power amplification and high \ac{OOBE}.

Significant effort has been made recently, therefore, to address these limitations.
In \cite{tao2024daft}, for instance, an access scheme was proposed in which the \ac{DFT} operation of  a spread-\ac{OFDM} system (DFT-s-OFDMA) was replaced by a \ac{DAFT}, yielding a spread-\ac{AFDM} access (DAFT-s-AFDMA) approach shown to have improved \ac{PAPR} properties.
In turn, in \cite{omar2020spectrum} a modified version of \ac{OCDM} dubbed c-\ac{OCDM} was presented, where spectral containment is improved by applying zero-padding in the frequency domain on the chirp matrices prior to modulation.
And given that \ac{AFDM} and \ac{OCDM} differ mainly on the choice of chirp parameters, the spectral containment approach of \cite{omar2020spectrum}, it is natural that a similar idea was presented for AFDM in \cite{savaux2024}, where a modified transceiver structure to reduce the overall computational complexity is also shown.

Approaching the problem from the perspective of \ac{OOBE}, on the other hand, \ac{FBMC} emerges as an alternative to design \ac{DD}-robust \ac{ISAC} waveforms, due to the per-subcarrier filtering process inherent of filter-bank methods. 
And while regular \ac{FBMC} systems are limited to real orthogonality and \ac{OQAM} modulation to fulfill the Balian-Low theorem, precoded versions such as those proposed in \cite{zakaria2012novel, nissel2018pruned, pereira2021novel, pereira2022generalized} allow the restoration of complex orthogonality, with additional low \ac{PAPR} features obtained via \ac{DFT}-pruning.

\Ac{AFBM}, first proposed in \cite{senger2025affinefilterbankmodulation}, is a novel multicarrier waveform designed to enable \ac{ISAC} in \ac{DD} channels while maintaining the low \ac{PAPR} and \ac{OOBE} characteristics of typical filter bank-based multicarrier systems.
However, the receiver design for \ac{AFBM} has not been addressed yet, and the existing approaches for \ac{AFDM} are not directly applicable due to the additional complexity introduced by the chirp-based subcarriers and the filtering process.
In this paper, we propose a low-complexity receiver structure for \ac{AFBM} based on the \ac{GaBP} framework, which is capable of performing detection of the transmitted symbols using only element-wise scalar operations, as demonstrated via simulation results.


\section{Signal Model}
\label{secSysModel}

Consider a point-to-point communication system with a single-antenna \ac{TX} and \ac{RX} operating in a \ac{DD} channel, using the proposed \ac{AFBM} waveform.
Following the typical filter bank-based multicarrier system structure, transmit symbols are mapped onto a grid with $L$ subcarriers and $K$ time indices. 
However, in order to maintain complex orthogonality, and since half of the system's $L$ subcarriers must be reserved as a guard interval to avoid filter interference, transmission is carried out at a double rate with respect to the conventional transmission seen in \ac{OFDM}/\ac{AFDM} systems, namely, every $L/2$ samples (see \cite{Ranasinghe_ICNC2025_oversampling} for more details on the oversampling effect tradeoff with \ac{AFDM}).

The considered filter bank-based system can be characterized as a block structure composed of $K$ symbols with a duration of $T/2$ seconds, where each symbol has $L$ subcarriers spaced by $F$ Hz.
Thus, in the TF domain we obtain a grid with $L$ points spaced by $F$ Hz along the frequency axis and $K$ points along the time axis with a space of $T/2$ seconds.
In this sense, the total bandwidth is defined as $B=LF$ and the total transmission time interval as $KT/2$.

\subsection{Transmit Signal}

Let $\mathbf{x} \in \mathcal{D}^{K\frac{L}{2} \times 1}$ denote the vector containing the transmit symbols, where $\mathcal{D}$ denotes an arbitrary modulation alphabet of size $|\mathcal{D}|$, such as \ac{QAM}.

The symbols in $\mathbf{x}$ are organized via insertion in the first and last $L/4$ positions in a matrix $\mathbf{A} \in \mathbb{C}^{L \times K}$ according to a pre-established transmission strategy in order to avoid interference from the filter bank, given by the relationship
\vspace{-0.5ex}
\begin{equation}
\label{eq:positions}
\mathbf{a} \triangleq \textrm{vec}(\mathbf{A}) = \bm{\Xi} \mathbf{x} \in \mathbb{C}^{LK \times 1},
\vspace{-0.5ex}
\end{equation}
where $\textrm{vec}(\cdot)$ denotes the column-wise vectorization operation and $\bm{\Xi} \in \mathbb{C}^{LK \times K\frac{L}{2}}$ is constructed as
\vspace{-0.5ex}
\begin{equation}
\label{eq:Xi}
\bm{\Xi} \triangleq \mathbf{I}_K \otimes \bar{\mathbf{\Xi}},
\vspace{-0.5ex}
\end{equation}
with $\bar{\mathbf{\Xi}} \in \mathbb{C}^{L \times \frac{L}{2}}$ defined as
\begin{equation}
\bar{\mathbf{\Xi}} \triangleq \begin{bmatrix}
\mathbf{I}_{L/4} & \mathbf{0}_{L/4} \\
\mathbf{0}_{L/2} & \mathbf{0}_{L/2} \\
\mathbf{0}_{L/4} & \mathbf{I}_{L/4}  
\end{bmatrix},
\end{equation}
where $\mathbf{0}_L$ denotes a full zero matrix of size $L$.

Then, the vectorized form of the DAFT-spread transmit signal $\mathbf{b} \in \mathbb{C}^{LK \times 1}$ whose matrix form is defined as $\mathbf{B} \in \mathbb{C}^{L \times K}$ before the prototype filter is given by
\begin{align}
\label{eq:precodede_tx}
\mathbf{b} & \triangleq \mathrm{vec}(\mathbf{B}) = \mathrm{vec}\big(\overbrace{\mathbf{W}_L\mathrm{diag} (\tilde{\mathbf{b}})}^{\triangleq \mathbf{C}_f \, \in \, \mathbb{C}^{L \times L}}\mathbf{A}\big) = \!\! \overbrace{\big(\mathbf{I}_K \otimes \mathbf{C}_f \big)}^{\triangleq \mathbf{C} \, \in \, \mathbb{C}^{LK \times LK}}\!\! \,\mathrm{vec}(\mathbf{A})\\[-3ex]
& =  \mathbf{C} \mathbf{a} = \mathbf{C} \bm{\Xi} \mathbf{x}. \nonumber
\end{align}
where $\mathbf{W}_{L} \in \mathbb{C}^{L \times L}$ represents the $L$-point DAFT matrix, $i.e.$
\vspace{-0.5ex}
\begin{equation}
\mathbf{W}_{L} = \mathbf{\Lambda}_{c_1,L}\mathbf{F}_{L}\mathbf{\Lambda}_{c_2,L},
\vspace{-0.5ex}
\end{equation}
with $\mathbf{\Lambda}_{c_i,L} = \text{diag}[e^{-j2\pi c_i (0)^2} , \dots, e^{-j2\pi c_i (L-1)^2}] \in \mathbb{C}^{L \times L}$ denoting an $L$-size diagonal chirp matrix with a central digital frequency of $c_i$, $\mathbf{F}_{L}$ a normalized $L$-point \ac{DFT} matrix and $\tilde{\mathbf{b}} \in \mathbb{C}^{L \times 1}$ the filter compensation vector. 

We emphasize that $\tilde{\mathbf{b}}$ is needed for transceiver complex orthogonality, as detailed in \cite[Eq.(12)]{senger2025affinefilterbankmodulation}.
From the above, the \ac{DAFT} $\mathbf{\tilde{W}}_{P} \in  \mathbb{C}^{L \times P}$ can be expressed as
\vspace{-0.5ex}
\begin{equation}
\mathbf{\tilde{W}}_{P} = \begin{bmatrix}
\mathbf{I}_L &  \mathbf{0}_{L\times (P-L)}  \end{bmatrix}
\mathbf{W}_P.
\label{dft_espalhada}
\vspace{-0.5ex}
\end{equation}

Afterwards, by using frequency domain zero padding, the output matrix $\mathbf{Q}_{P}$ for a given block can be generated as
\vspace{-0.5ex}
\begin{equation}
\mathbf{Q}_{P} = \mathbf{F}_{N}^{H}\mathbf{T}\mathbf{F}_P \mathbf{\tilde{W}}^H_{P},
\label{eq:Q_P}
\vspace{-0.5ex}
\end{equation}
where $\mathbf{T}\triangleq \begin{bmatrix} \mathbf{I}_{P,l}^{T} & \mathbf{0}_{(N-P)\times P} & \mathbf{I}_{P,u}^{T} \end{bmatrix}$,
with $\mathbf{T}^{T}\mathbf{T} = \mathbf{I}_P$, is an $N \times P$ matrix, with $\mathbf{I}_{P,u}$ and $\mathbf{I}_{P,l}$ denoting the first and last $P/2$ columns of the $P \times P$ identity matrix $\mathbf{I}_P$, respectively.

Considering the transmission of $K$ blocks, the block matrix $\mathbf{Q} \in  \mathbb{C}^{NK\times LK}$ is defined as
\begin{equation}
\mathbf{Q} =  \mathbf{I}_{K} \otimes \mathbf{Q}_{P} , 
\vspace{-0.5ex}
\end{equation}
where $\otimes$ refers to the Kronecker product which maps $\mathbf{Q}_{P}$ to the correct time positions.
\newpage

Then, the transmitted data can be obtained by convolving the output with the prototype filter impulse response through a Toeplitz filter matrix.
Let us consider the diagonal matrix $\mathbf{G}_p$ corresponding to the filter coefficients,  that is, $\mathbf{G}_p =$ diag$(\mathbf{g}_p) \in  \mathbb{R}^{N/2 \times N/2}$ for $p = 0,1,2,...,2 O-1$, where $\mathbf{g}_{p}$ is given by  $\mathbf{g}_p = [g[pN/2],g[pN/2+1],   \ldots , g[pN/2+N/2-1]]$ and $\mathbf{g}$ is the prototype filter of length $ON$, where $O$ is the filter overlap factor.
Thus, the block Toeplitz filter matrix $\mathbf{G} \in  \mathbb{R}^{ON + (K-1)N/2 \times NK}$ can be generated from the vector $[ \mathbf{G}_0  \hspace{0.1cm}  \mathbf{G}_1   \hspace{0.1cm} \mathbf{G}_2 \ldots \mathbf{G}_{2O -1}]^T$, as detailed in \cite[Eq.(7)]{senger2025affinefilterbankmodulation}.
%

Then, the full \ac{AFBM} transmit signal in the \ac{TD} can be efficiently described in terms of the filter matrix $\mathbf{G}$, modified \ac{IDAFT} matrix $\mathbf{Q}_P$, and the compensation matrix $\mathbf{C}_f$ leveraging identities involving Kronecker products as
\vspace{-1ex}
\begin{align}
\label{eq:td_tx_signal}
\mathbf{s} & = \mathbf{G} \mathbf{Q} \mathbf{C} \mathbf{a}  = \mathbf{G} \big(\mathbf{I}_{K} \otimes \mathbf{Q}_{P}\big) \cdot \big(\mathbf{I}_K \otimes \mathbf{C}_f \big) \mathbf{a} \in \mathbb{C}^{M \times 1} \nonumber \\
& = \mathbf{G} \big(\mathbf{I}_{K} \otimes \mathbf{Q}_{P} \mathbf{C}_f \big) \bm{\Xi} \mathbf{x},
\end{align}
where $M \triangleq ON + \frac{N}{2}(K-1)$, and the symbols are delayed from each other every $N/2$ samples, as clarified by the blocks $\mathbf{0}_{N/2}$ in $\mathbf{G}$ \cite{pereira2022generalized}.

From the above, the proposed structure is an affine-precoded filter bank scheme can be implemented efficiently using a \ac{PPN} together with an \ac{IFFT}.
Another interpretation is that in the proposed waveform the standard AFDM sinc chirp subcarriers were replaced with chirp filtered ones.
In addition, the length $P$ of the \ac{IDAFT} at the transmitter must be greater than $L$ to prevent the precoding stage \ac{DAFT} from being nullified by the \ac{IDAFT} of the filter bank structure.
Finally, $P$ must be smaller than the size of the filter bank $N$ so that the chirps are sampled at a rate lower than Nyquist's, allowing frequency containment.

\subsection{Received Signal}
\label{subsec:received_signal}

The signal vector $\mathbf{s}$ defined in equation \eqref{eq:td_tx_signal} is transmitted over a time-varying multipath channel, i.e., the doubly-dispersive channel, described concisely by the circular convolutional matrix form $\mathbf{H} \in  \mathbb{C}^{M \times M}$ \cite{nissel2018pruned}. Consequently, the received signal $\mathbf{r} \in  \mathbb{C}^{M \times 1}$  is expressed as
\vspace{-0.5ex}
\begin{equation}
\mathbf{r} \triangleq 
\Big(\sum_{r=1}^{R}h_r \mathbf{\Phi}_r\mathbf{Z}^{f_r} \mathbf{\Pi}^{\ell_r}\Big)
\mathbf{G} \Big(\mathbf{I}_{K} \otimes \mathbf{Q}_{P} \mathbf{C}_f\Big) \bm{\Xi} \mathbf{x} + \mathbf{n},
\vspace{-0.5ex}
\end{equation}
where $\mathbf{n} \in \mathbb{C}^{M \times 1}$ represents \ac{AWGN} samples with zero mean and power $\sigma_n^2$, while $\mathbf{H} \triangleq \sum_{r=1}^{R}h_r \mathbf{\Phi}_{r} \mathbf{Z}^{f_r} \mathbf{\Pi}^{\ell_r} \in \mathbb{C}^{M \times M}$, parametrized by the complex channel fading coefficient $h_r \in \mathbb{C}$, is the doubly-dispersive wireless channel described by $R$ resolvable propagation paths, and it is implicitly assumed that each $r$-th scattering path induces a delay $\tau_r \in [0, \tau^\mathrm{max}]$ and Doppler shift $\nu_r \in [-\nu^\mathrm{max}, +\nu^\mathrm{max}]$ to the received signal, where $\ell_r \triangleq \lfloor \frac{\tau_r}{T_\mathrm{s}} \rceil \in \mathbb{N}_0$ and $f_r \triangleq \frac{N\nu_r}{f_\mathrm{s}} \in \mathbb{R}$ are the corresponding normalized integer path delay and normalized digital Doppler shift, with sampling frequency $f_\mathrm{s} \triangleq \frac{1}{T_\mathrm{s}}$. 

As elaborated in \cite{Rou_SPM_2024}, each $r$-th path of the full circular convolutional matrix $\mathbf{H}$, is parametrized by the complex channel fading coefficient $h_r \in \mathbb{C}$, the diagonal prefix phase matrix $\boldsymbol{\Phi}_r  \!\in\! \mathbb{C}^{M \times M}$ with the IDAFT-based chirp-cyclic prefix phase function $\phi(m) \triangleq  c_1(M^2 - 2Mm) $, given by
\vspace{-0.5ex}
\begin{equation}
\color{black}
\!\!\!\!\!\mathbf{\Phi}_{r} \!\triangleq\! \mathrm{diag}\big[e^{-j2\pi \cdot \phi(\ell_r)}, \ldots, e^{-j2\pi \cdot\phi(1)}, 1, \ldots, 1\big] \!\in\! \mathbb{C}^{M \times M}\!,\!\!\!
\label{eq:CCP_phase_matrix}
\end{equation}
the diagonal roots-of-unity matrix $\mathbf{Z} \!\in\! \mathbb{C}^{M \times M}$ given by 
\vspace{-1ex}
\begin{equation}
    \vspace{-1ex}
\mathbf{Z} \triangleq \mathrm{diag}\big[e^{-j2\pi\frac{(0)}{M}}, \,\ldots\,, e^{-j2\pi\frac{(M\!-\!1)}{M}}\big] \in \mathbb{C}^{M \times M},
\label{eq:Z_matrix}
\end{equation}
which is taken to the $f_r$-th power, and the right-multiplying circular left-shift matrix $\mathbf{\Pi} \in \mathbb{C}^{M \times M}$.

%
%

Next, the received signal is demodulated by $(\mathbf{G}\mathbf{Q})^H$, generating $\mathbf{\tilde{a}}$ $\in \mathbb{C}^{LK \times 1}$.
Then (and considering all time instants), the detected symbols represented by $\mathbf{\tilde{B}} \in C^{L \times K}$ are obtained through the \ac{IDAFT} combined with the compensation stage
\vspace{-1ex}
\begin{equation}
    \vspace{-1ex}
\mathbf{\tilde{B}} = \mathbf{W}^H_L \textrm{diag}\{\tilde{\mathbf{b}}\} \mathbf{\tilde{A}},
\label{12ghtj}
\end{equation}
where $\mathbf{\tilde{A}} = [\mathbf{\tilde{a}}_0;\mathbf{\tilde{a}}_1;\hdots;\mathbf{\tilde{a}}_{K-1}] \in \mathbb{C}^{L \times K}$.

To obtain the estimated symbols, the $L/2$ intermediate values between the first and last $L/4$ symbols are discarded, since no data is at those positions in $\mathbf{A}$.
For symmetry reasons, the compensation stage is used in both the receiver and transmitter, which can be applied only to one side effectively.

Concatenating all these effects, the final received signal $\mathbf{y} \in \mathbb{C}^{K\frac{L}{2} \times 1}$ after demodulation without noise can be expressed as
\vspace{-1ex}
\begin{equation}
    \vspace{-1ex}
\label{eq:final_IO}
\mathbf{y} \triangleq \bm{\Xi}^H \big(\mathbf{I}_{K} \otimes \mathbf{C}_f^H \mathbf{Q}_{P}^H \big) \mathbf{G}^H \mathbf{H} \mathbf{G} \big(\mathbf{I}_{K} \otimes \mathbf{Q}_{P} \mathbf{C}_f\big) \bm{\Xi} \mathbf{x},
\end{equation}
such that an end-to-end effective channel matrix can be extracted, which is given by
%
\begin{equation}
\mathbf{H}_\text{eff} \triangleq \bm{\Xi}^H \big(\mathbf{I}_{K} \otimes \mathbf{C}_f^H \mathbf{Q}_{P}^H \big) \overbrace{\mathbf{G}^H \mathbf{H} \mathbf{G} \big(\mathbf{I}_{K} \otimes \mathbf{Q}_{P} \mathbf{C}_f\big) \bm{\Xi}}^{\bar{\mathbf{H}} \in \mathbb{C}^{NK \times K\frac{L}{2}}}.
\label{eq:Heff}
\end{equation}

\section{Proposed GaBP-based Receiver Design}
\label{secReceiver}

From a receiver design viewpoint, we aim to estimate the transmit signal $\mathbf{x}$, under the assumption that the filtered \ac{TD} channel matrix $\bar{\mathbf{H}} \in \mathbb{C}^{NK \times K\frac{L}{2}}$ is known\footnote{The final \ac{I/O} relationship portrayed in equation \eqref{eq:final_IO} is not used since the Gram matrix of $\mathbf{H}_\text{eff}$ does not approach a clear diagonal.}, such that in order to derive a \ac{GaBP}-based detector for the \ac{AFBM} waveforms, we consider the \ac{I/O} relationship given by
\vspace{-1ex}
\begin{equation}
\label{General_I/O_arbitrary}
\bar{\mathbf{r}} = \bar{\mathbf{H}}  \mathbf{x} + \bar{\mathbf{w}}.
\vspace{-1ex}
\end{equation}

For comparison, the equivalent high-complexity \ac{LMMSE} can be expressed as
\vspace{-1ex}
\begin{equation}
\label{eq:LMMSE}
\hat{\mathbf{x}}_\text{LMMSE} = \left( \bar{\mathbf{H}}^H \bar{\mathbf{H}} + \sigma^2_n \mathbf{I}_{\bar{M}} \right)^{-1} \bar{\mathbf{H}}^H \bar{\mathbf{r}}. 
\vspace{-1ex}
\end{equation}

Note that as a consequence of this approach, no additional demodulation stage is required after the effects of $\mathbf{G}$ are reversed for data detection.

Setting $\bar{N} \triangleq NK$ and $\bar{M} \triangleq K\frac{L}{2}$ with $\bar{n} \triangleq \{1,\dots,\bar{N}\}$ and $\bar{m} \triangleq \{1,\dots,\bar{M}\}$, the element-wise relationship corresponding to equation \eqref{General_I/O_arbitrary} is given by
\vspace{-1ex}
\begin{equation}
\label{General_I/O_arbitrary_elementwise}
\bar{r}_{\bar{n}} = \sum_{\bar{m}=1}^{\bar{M}} \bar{h}_{\bar{n},\bar{m}} x_{\bar{m}} + \bar{w}_{\bar{n}}, 
\vspace{-1ex}
\end{equation}
such that the soft replica of the $\bar{m}$-th communication symbol associated with the $\bar{n}$-th receive signal $\bar{r}_{\bar{n}}$, computed at the $i$-th iteration of a message-passing algorithm can be denoted by $\hat{x}_{\bar{n},\bar{m}}^{(i)}$, with the corresponding \ac{MSE} of these estimates computed for the $i$-th iteration given by
\vspace{-0.5ex}
\begin{equation}
\hat{\sigma}^{2(i)}_{x:{\bar{n},\bar{m}}} \triangleq \mathbb{E}_{x} \big[ | x - \hat{x}_{\bar{n},\bar{m}}^{(i-1)} |^2 \big]= E_\mathrm{S} - |\hat{x}_{\bar{n},\bar{m}}^{(i-1)}|^2, \forall (\bar{n},\bar{m}),
\label{eq:MSE_d_k}
\vspace{-0.5ex}
\end{equation}
where $\mathbb{E}_{x}$ denotes expectation over all possible symbols $x$.

The \ac{GaBP} receiver for such a setup consists of three major stages described below.

\subsubsection{Soft Interference Cancellation} At a given $i$-th iteration of the \ac{sIC} stage of the algorithm, the soft replicas $\hat{x}_{\bar{n},\bar{m}}^{(i-1)}$ from a previous iteration are used to calculate the data-centric \ac{sIC} signals $\tilde{r}_{x:\bar{n},\bar{m}}^{(i)}$.

Exploiting equation \eqref{General_I/O_arbitrary_elementwise}, the \ac{sIC} signals are given by
\vspace{-1ex}
\begin{equation}
\label{eq:d_soft_IC}
\tilde{r}_{x:\bar{n},\bar{m}}^{(i)}\!\! = \!\bar{r}_{\bar{n}} -\! \sum_{e \neq \bar{m}}\!\! h_{\bar{n},e} \hat{x}_{\bar{n},e}^{(i)}, 
= h_{\bar{n},\bar{m}} x_{\bar{m}} + \!\overbrace{\sum_{e \neq \bar{m}}\!\! h_{\bar{n},e}(x_e\!\! -\! \hat{x}_{\bar{n},e}^{(i)})\! +\! \bar{w}_{\bar{n}}}^\text{interference + noise term},\nonumber
\end{equation}

Using the \ac{SGA}, the interference and noise terms highlighted above can be approximated as Gaussian noise, such that the conditional \acp{PDF} of the \ac{sIC} signals become
\vspace{-0.5ex}
\begin{equation}
\label{eq:cond_PDF_d}
\!\!p_{\tilde{\mathrm{r}}_{\mathrm{x}:\bar{n},\bar{m}}^{(i)} \mid \mathrm{x}_{\bar{m}}}(\tilde{r}_{x:\bar{n},\bar{m}}^{(i)}|x_{\bar{m}}) \propto \mathrm{exp}\bigg[ -\frac{|\tilde{r}_{x:\bar{n},\bar{m}}^{(i)}\! -\! h_{\bar{n},\bar{m}} x_{\bar{m}}|^2}{\tilde{\sigma}_{x:\bar{n},\bar{m}}^{2(i)}} \bigg]\!,
\vspace{-0.5ex}
\end{equation}
with their conditional variances expressed as
\vspace{-0.5ex}
\begin{equation}
\label{eq:soft_IC_var_d}
\tilde{\sigma}_{x:\bar{n},\bar{m}}^{2(i)} = \sum_{e \neq \bar{m}} \left|h_{\bar{n},e}\right|^2 \hat{\sigma}^{2(i)}_{x:{\bar{n},e}} + \sigma^2_n.
\end{equation}


\subsubsection{Belief Generation} In the belief generation stage of the algorithm the \ac{SGA} is exploited under the assumptions that $\bar{N}$ is a sufficiently large number, and that the individual estimation errors in $\hat{x}_{\bar{n},\bar{m}}^{(i-1)}$ are independent, in order to generate initial estimates (aka beliefs) for all the data symbols.

As a consequence of the \ac{SGA} and with the conditional \acp{PDF} of equation \eqref{eq:cond_PDF_d}, the following extrinsic \acp{PDF}
\begin{equation}
\label{eq:extrinsic_PDF_d}
\prod_{e \neq \bar{n}} p_{\tilde{\mathrm{r}}_{\mathrm{x}:e,\bar{m}}^{(i)} \mid \mathrm{x}_{\bar{m}}}(\tilde{r}_{x:e,\bar{m}}^{(i)}|x_{\bar{m}}) \propto \mathrm{exp}\bigg[ - \frac{(x_{\bar{m}} - \bar{x}_{\bar{n},\bar{m}}^{(i)})^2}{\bar{\sigma}_{x:\bar{n},\bar{m}}^{2(i)}} \bigg],
\end{equation}
are obtained, where the corresponding extrinsic means and variances are respectively defined as
%
%
\begin{equation}
\label{eq:extrinsic_mean_d}
\bar{x}_{\bar{n},\bar{m}}^{(i)} = \bar{\sigma}_{x:\bar{n},\bar{m}}^{(i)} \sum_{e \neq \bar{n}} \frac{h^*_{e,\bar{m}} \tilde{r}_{x:e,\bar{m}}^{(i)}}{ \tilde{\sigma}_{x:e,\bar{m}}^{2(i)}} \;\text{and}\;
%
\bar{\sigma}_{x:\bar{n},\bar{m}}^{2(i)} \!=\! \bigg(\! \sum_{e \neq \bar{n}} \frac{|h_{e,\bar{m}}|^2}{\tilde{\sigma}_{x:e,\bar{m}}^{2(i)}} \bigg)^{\!\!\!-1}\!\!\!\!,
\end{equation}
with $h^*_{e,\bar{m}}$ denoting the complex conjugate of $h_{e,\bar{m}}$.

\subsubsection{Soft Replica Generation} Finally, the soft replica generation stage consists of denoising the previously computed beliefs under a Bayes-optimal rule, to obtain the final estimates for the desired variables.
For \ac{QPSK} modulation\footnote{We consider \ac{QPSK} for simplicity, but \ac{wlg}, since denoisers for other modulation schemes can also be designed \cite{TakahashiTCOM2019}.}, the Bayes-optimal denoiser is given by
\vspace{-1ex}
\begin{equation}
\hat{x}_{\bar{n},\bar{m}}^{(i)}\! =\! c_x \bigg(\! \text{tanh}\!\bigg[ 2c_d \frac{\Re{\bar{x}_{\bar{n},\bar{m}}^{(i)}}}{\bar{\sigma}_{x:{\bar{n},\bar{m}}}^{2(i)}} \bigg]\!\! +\! \jmath \text{tanh}\!\bigg[ 2c_d \frac{\Im{\bar{x}_{\bar{n},\bar{k}}^{(i)}}}{\bar{\sigma}_{{x}:{\bar{n},\bar{m}}}^{2(i)}} \bigg]\!\bigg),\!\!
\label{eq:QPSK_denoiser}
\end{equation}
where $c_x \triangleq \sqrt{E_\mathrm{S}/2}$ denotes the magnitude of the real and imaginary parts of the explicitly chosen \ac{QPSK} symbols, with its corresponding variance updated as in equation \eqref{eq:MSE_d_k}.

After obtaining $\hat{x}_{\bar{n},\bar{m}}^{(i)}$ as per equation \eqref{eq:QPSK_denoiser}, the final outputs are computed by damping the results to prevent convergence to local minima due to incorrect hard-decision replicas \cite{Su_TSP_2015}.
Letting the damping factor be $0 < \beta_x < 1$ yields
\vspace{-1ex}
\begin{subequations}
\begin{equation}
\vspace{-1ex}
\label{eq:d_damped}
\hat{x}_{\bar{n},\bar{m}}^{(i)} = \beta_x \hat{x}_{\bar{n},\bar{m}}^{(i)} + (1 - \beta_x) \hat{x}_{\bar{n},\bar{m}}^{(i-1)}.
\end{equation}
\begin{equation}
\vspace{-1ex}
\label{eq:MSE_d_m_damped}
\hat{\sigma}^{2(i)}_{x:{\bar{n},\bar{m}}} = \beta_x (E_\mathrm{S} - |\hat{x}_{\bar{n},\bar{m}}^{(i-1)}|^2) + (1-\beta_x) \hat{\sigma}_{x:{\bar{n},\bar{m}}}^{2(i-1)},
\end{equation}
\end{subequations}

Finally, as a result of the conflicting dimensions, the consensus update of the estimates can be obtained as
\vspace{-1ex}
\begin{equation}
\vspace{-1ex}
\label{eq:d_hat_final_est}
\hat{x}_{\bar{m}} = \bigg( \sum_{\bar{n}=1}^{\bar{N}} \frac{|h_{\bar{n},\bar{m}}|^2}{\tilde{\sigma}_{x:\bar{n},\bar{m}}^{2(i_\text{max})}} \bigg)^{\!\!\!-1} \! \! \bigg( \sum_{\bar{n}=1}^{\bar{N}} \frac{h^*_{\bar{n},\bar{m}} \tilde{r}_{x:\bar{n},\bar{m}}^{(i_\text{max})}}{ \tilde{\sigma}_{x:\bar{n},\bar{m}}^{2(i_\text{max})}} \bigg).
\end{equation}


\vspace{1ex}
\section{Performance Analysis}
\label{secSymResults}


The complexity of the proposed \ac{GaBP} detection algorithm is linear on the number of element-wise operations, and its per-iteration computational complexity is given by $\mathcal{O}(\bar{N}\bar{M})$. 
Notice that this complexity is much lesser than that of typical detection methods such as the \ac{LMMSE}, which is $\mathcal{O}(\bar{M}^3)$ due to the costly matrix inversion involved.


For the simulations, the total number of subcarriers $L$ is 128, the chirp size $P$ is 256 and the filter bank \ac{DFT} size $N$ is 256.
Each transmission was composed of $K = 8$ symbols, with a carrier frequency $f_c$ of 4 GHz.
The channel is a doubly dispersive one, with three resolvable paths and corresponding normalized delays and digital Doppler shifts.
We recall that the chirp frequencies for each (I)DAFT were chosen to uphold the orthogonality condition \cite{Rou_SPM_2024} $2(f^{\text{max}} + \xi)(\ell^{\text{max}}+1) + \ell^{\text{max}} \leq P$, where  $f^{\text{max}}$ and $\ell^{\text{max}}$ are, respectively, the maximum normalized digital Doppler shift and delay of the channel and $\xi \in \mathbb{N}_0$ is a free parameter determining the so-called guard width, denoting the number of additional guard elements around the diagonals to anticipate for Doppler-domain interference.

\begin{figure}[H]
\centering
\includegraphics[width=0.95\columnwidth]{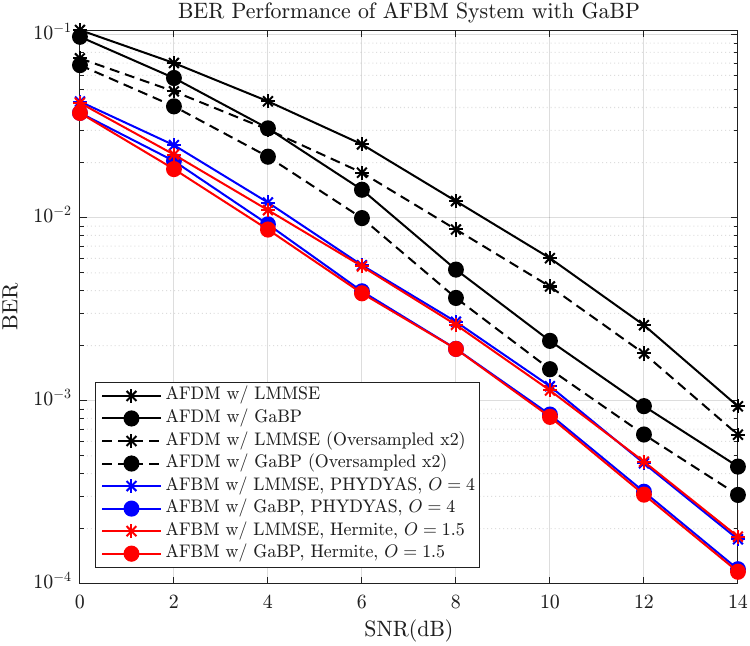}%
\vspace{-2ex}
\caption{\ac{BER} performance of the proposed \ac{GaBP} technique for both the \ac{AFDM} and \ac{AFBM} waveforms with Hermite and PHYDYAS prototype filters.}
\label{fig:afdm_ber}
\vspace{1ex}
\includegraphics[width=0.95\columnwidth]{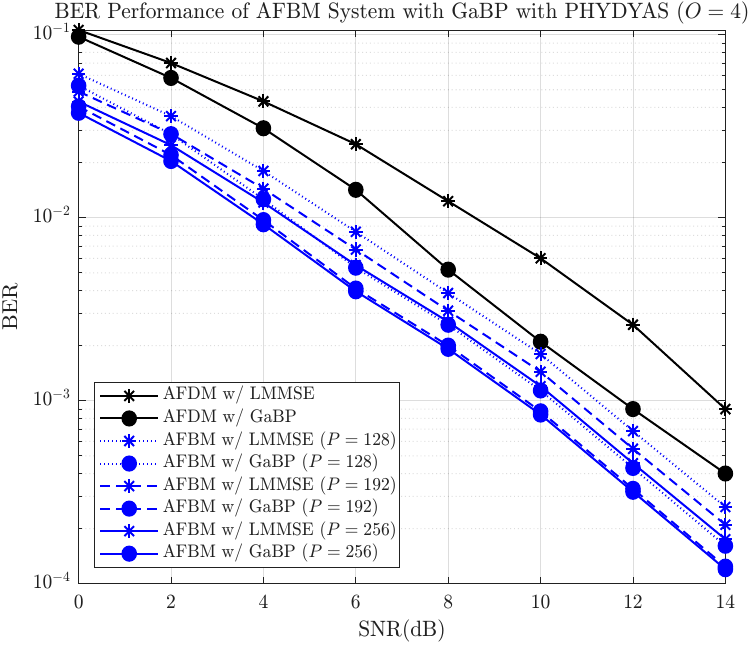}%
\vspace{-2ex}
\caption{\ac{BER} performance of the proposed \ac{GaBP} technique for both the \ac{AFDM} with PHYDYAS prototype filters for varying $P$.}
\label{fig:afdm_ber_p}
\vspace{1ex}
\includegraphics[width=0.95\columnwidth]{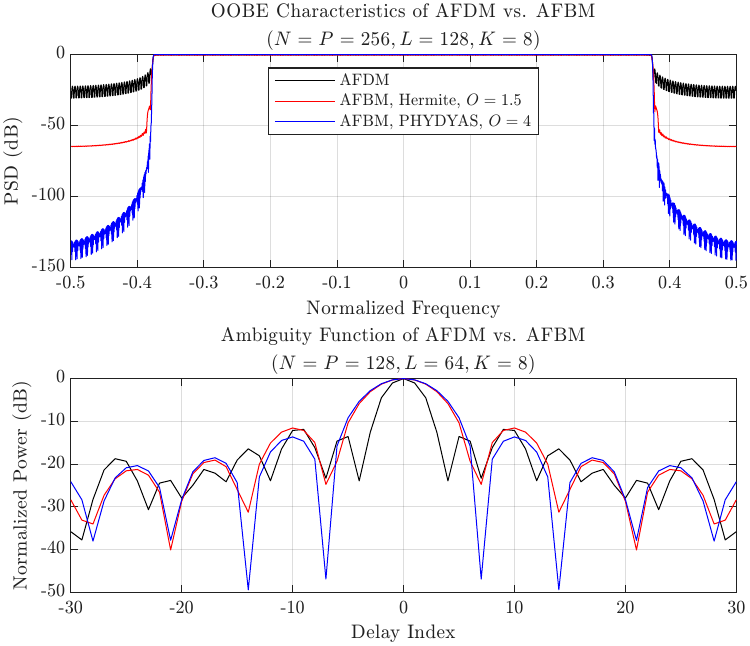}%
\vspace{-2ex}
\caption{\ac{OOBE} and Abmibuity function Performance of \ac{AFDM} and \ac{AFBM} with the Hermite and PHYDYAS prototype filters.}
\label{fig:afdm_oob}
\vspace{-2ex}
\end{figure}

Figures \ref{fig:afdm_ber} and \ref{fig:afdm_ber_p} presents \ac{BER} results from the considered systems.
We remark that as opposed to the \ac{TD} relationship in equation \eqref{General_I/O_arbitrary} considered for the proposed \ac{AFBM} scheme, the typical \ac{DAF} domain \ac{I/O} presented in \cite{RanasingheTWC2025} is used for \ac{AFDM} since the effective Gram matrix structures are almost identical.
It can be seen that, the proposed \ac{GaBP}-based \ac{AFBM} scheme outperforms the regular \ac{AFDM} scheme by approximately 2 dB at a \ac{BER} of $10^{-3}$, which is expected due to the better spectral localization and the reduced interference in the proposed scheme, even with varing $P$.
Next, the \ac{OOBE} and ambiguity functions of the considered schemes are shown in Figure~\ref{fig:afdm_oob}.
In this scenario, two prototype filters were considered for \ac{AFBM}: the truncated Hermite  (with $O = 1.5$) and the full PHYDYAS (with $O = 4$).
Due to the usage of well-localized filters instead of the rectangular window there is a significant advantage in \ac{OOBE} of the \ac{AFBM} system with respect to regular \ac{AFDM}.
Finally, the ambiguity function of the \ac{AFBM} system is also very similar that of the \ac{AFDM} system, with a slightly better sidelobe supression seen with the PHYDYAS prototype filter.

\balance
\newpage
\selectlanguage{english}
\bibliographystyle{IEEEtran}
\bibliography{references.bib}

\end{document}